\documentclass[aps,prl,preprint,groupedaddress]{revtex4}

\begin{document}

\title{Dirac Quantization of the Pais-Uhlenbeck Fourth Order Oscillator}

\author{Philip D. Mannheim}
\email[]{mannheim@uconnvm.uconn.edu}
\affiliation{Department of Physics, University of Connecticut,
Storrs, CT 06269}

\author{Aharon Davidson}
\email[]{davidson@bgumail.bgu.ac.il}
\affiliation{Physics Department, Ben Gurion University of the Negev,
Beer-Sheva 84105, Israel}

\date{January 16, 2005}

\begin{abstract}

As a model, the Pais-Uhlenbeck fourth order oscillator with equation of
motion $$\frac{d^4q}{dt^4}+(\omega_1^2+\omega_2^2)\frac{d^2q}{dt^2} 
+\omega_1^2\omega_2^2 q=0$$ is a quantum-mechanical prototype of a field
theory containing both second and fourth order derivative terms. With
its dynamical degrees of freedom obeying constraints due to the presence
of higher order time derivatives, the model cannot be quantized
canonically. We thus quantize it using the method of Dirac constraints
to construct the correct quantum-mechanical Hamiltonian for the system,
and find that the Hamiltonian diagonalizes in the positive and negative
norm states that are characteristic of higher derivative field theories.
However, we also find that the oscillator commutation relations become
singular in the $\omega_1 \rightarrow \omega_2$ limit, a limit which
corresponds to a prototype of a pure fourth order theory. Thus the
particle content of the $\omega_1 =\omega_2$ theory cannot be inferred
from that of the $\omega_1 \neq \omega_2$ theory; and in fact in the
$\omega_1 \rightarrow \omega_2$ limit we find that all of the
$\omega_1 \neq \omega_2$ negative norm states move off shell, with 
the spectrum of asymptotic in and out states of the equal frequency theory
being found to be completely devoid of states with either negative energy or
negative norm. As a byproduct of our work we find a Pais-Uhlenbeck analog
of the zero energy theorem of Boulware, Horowitz and Strominger, and show
how in the equal frequency Pais-Uhlenbeck theory the theorem can be
transformed into a positive energy theorem instead.  

\end{abstract}

\pacs{}

\maketitle

\section{Introduction}

While attention in physics has by and large concentrated on second
order equations of motion, nonetheless, from time to time there has also
been some interest in higher derivative theories, with a typical such
higher order equation of motion being the scalar field equation of
motion
\begin{equation}
(\partial_0^2-\nabla^2)(\partial_0^2-\nabla^2+M^2)\phi(\bar{x},t)=0~~,
\label{1}
\end{equation}
a wave equation which is based on both second and fourth order
derivatives of the field. With the propagator associated with Eq.
(\ref{1}) being given by
\begin{equation}
D(k^2,M^2)= \frac{1}{k^2(k^2+M^2)}=\frac{1}{M^2}\left[\frac{1}{k^2}
-\frac{1}{k^2+M^2}\right]
\label{4}
\end{equation}
in momentum space, the expectation of canonical reasoning (see e.g.
\cite{Stelle1977} for a canonical study of theories of such second plus
fourth order type) is that a quantization of the theory associated with
Eq. (\ref{1}) would, for $M^2 \neq 0$,  lead to a $1/k^2 -1/(k^2+M^2)$
dipole spectrum consisting of two species of particles, one possessing
positive signature and the other negative or ghost signature. And
indeed, part of the appeal of such propagators is that precisely because
of this ghost signature, the propagator has much better behavior in the
ultraviolet than a standard second order $1/k^2$ propagator, to thus
enable this higher order theory to naturally address renormalization
issues such as those associated with elementary particle self-energies
or with quantum gravitational fluctuations.

While a similar conclusion regarding the presence of ghosts might
be anticipated to apply to pure fourth order theories as well (viz.
the $M^2=0$ limit of Eq. (\ref{1})),\cite{footnote0} as we see from
the form of Eq. (\ref{4}), the $1/M^2$ prefactor multiplying the
$1/k^2-1/(k^2+M^2)$ dipole term is singular, with it thus not being
reliable to try to infer the structure of the $M^2=0$ spectrum from
that associated with the $M^2 \neq 0$ one. In fact, it is actually not
even reliable to try to infer the spectrum associated with
Eq. (\ref{1}) via canonical reasoning at all, since even when $M^2
\neq 0$, the theory is a constrained one due to the presence of
higher order time derivatives; and thus before drawing any
conclusions at all, one must first quantize the theory in a way which
fully takes these constraints into account, and only then identify
the particle spectrum. To address this issue we shall thus effect a
full Dirac constraint quantization
\cite{Dirac1964} of a quantum-mechanical prototype of Eq. (\ref{1}),
something which as far as we know has not previously been carried out
in the literature. Specifically, we shall study a restricted version of
Eq. (\ref{1})  in which we specialize to field configurations of the
form $\phi(\bar{x},t)=q(t)e^{i\bar{k}\cdot\bar{x}}$, configurations in
which  Eq. (\ref{1}) then reduces
\begin{equation}
\frac{d^4q}{dt^4} +(\omega_1^2+\omega_2^2)\frac{d^2q}{dt^2} 
+\omega_1^2\omega_2^2 q=0~~,
\label{2}
\end{equation}
where
\begin{equation}
\omega_1^2+\omega_2^2=2\bar{k}^2+M^2~~,~~\omega_1^2\omega_2^2
=\bar{k}^4+\bar{k}^2M^2~~,
\label{3}
\end{equation}
with Eq. (\ref{3}) itself reducing to the equal frequency
$\omega_1=\omega_2$ when $M=0$. As such, for general
$\omega_1\neq \omega_2$ Eq. (\ref{2}) thus serves as a
quantum-mechanical prototype of a field theory based on second order
plus fourth order derivatives of the field, while becoming a
prototype of a pure fourth order theory in the equal frequency limit.
As well as being a model which we can actually solve analytically,
the quantum-mechanical prototype encapsulates many of the general
issues associated with the quantization of the full field theory,
since it is the time derivatives of the fields rather then their
spatial derivatives which are of relevance for constructing
conjugates of the fields. With the equation of motion given in Eq.
(\ref{2}) being derivable from the Pais-Uhlenbeck Lagrangian 
\cite{Pais1950} 
\begin{equation}
L=\frac{\gamma}{2}\left[\ddot{q}^2
-(\omega_1^2+\omega_2^2)\dot{q}^2+\omega_1^2\omega_2^2q^2\right]
\label{5}
\end{equation}
($\gamma$ is a constant), we shall thus quantize the theory
starting from this Lagrangian.\cite{footnote1} This will allow us to
establish that the properly constructed $\omega_1 \neq \omega_2$
Hamiltonian does indeed possess eigenstates of normal and ghost signature
(states which themselves are then explicitly constructed), while also
allowing us to monitor the $\omega_1 \rightarrow \omega_2$  limit, a
limit, which like the above $M^2 \rightarrow 0$ limit, will be found
to be highly singular.\cite{footnote2}

\section{Classical Dirac Hamiltonian}

For a quantization of the theory associated with the Lagrangian of Eq.
(\ref{5}) we would like to treat $q$ and $\dot{q}$ as independent
coordinates, but cannot immediately do so since if $\dot{q}$ is to be
an independent coordinate, we could not then use $\partial L/ \partial
\dot{q}$ as the canonical conjugate of $q$. To obtain a form
which would be appropriate for quantization we therefore introduce a
new variable $x(t)$ to replace $\dot{q}$, and compensate for doing so by
additionally introducing a Lagrange multiplier $\lambda(t)$ and a
substitute Lagrangian
\begin{equation}
L=\frac{\gamma}{2}\left[\dot{x}^2-(\omega_1^2+\omega_2^2)x^2+
\omega_1^2\omega_2^2q^2\right]+\lambda(\dot{q}-x)~~,
\label{6}
\end{equation}
with the Dirac constraint method assuring us that the Hamiltonian which
is to ultimately emerge will then be independent of the Lagrange
multiplier. With the Lagrangian of Eq. (\ref{6}) possessing three
coordinate variables, $q$, $x$ and $\lambda$, we must introduce the
three canonical momenta, $p_x$, $p_q$ and $p_{\lambda}$, conjugates
which for the Lagrangian of Eq. (\ref{6}) evaluate to
\begin{equation}
p_x=\frac{\partial L}{\partial
\dot{x}} =\gamma \dot{x}~~,~~ p_q=\frac{\partial L}{\partial
\dot{q}}=\lambda~~,~~
p_{\lambda}=\frac{\partial L}{\partial \dot{\lambda}}=0~~.
\label{7}
\end{equation}
To implement the Dirac method, as well as use these relations to
construct the Legendre transform of the Lagrangian of Eq. (\ref{6}),
viz.
\begin{equation}
H_L=p_x\dot{x}+p_q\dot{q}+p_{\lambda}\dot{\lambda}-L~~,
\label{9}
\end{equation}
we also introduce two primary constraint functions  for the canonical
momenta which involve either the Lagrange multiplier or its conjugate,
viz.
\begin{equation}
\phi_1=p_q-\lambda~~,~~ \phi_2=p_{\lambda}~~.
\label{8}
\end{equation}
And then, rather than use $H_L$, the Dirac prescription is to instead
use
\begin{equation}
H_1=H_L+u_1\phi_1+u_2\phi_2
\label{10}
\end{equation}
as Hamiltonian, where $H_1$ is thus given by
\begin{equation}
H_1=\frac{p_x^2}{2\gamma}+\frac{\gamma}{2}(\omega_1^2+\omega_2^2)x^2
-\frac{\gamma}{2}\omega_1^2\omega_2^2q^2+\lambda x
+u_1(p_q-\lambda)+u_2p_{\lambda}~~.
\label{11}
\end{equation}
For this theory we define generalized Poisson brackets of the form
\begin{equation}
\{A,B\}=\frac{\partial A}{\partial x}\frac{\partial B}{\partial p_x}
-\frac{\partial A}{\partial p_x}\frac{\partial B}{\partial x}
+\frac{\partial A}{\partial q}\frac{\partial B}{\partial p_q}
-\frac{\partial A}{\partial p_q}\frac{\partial B}{\partial q}
+\frac{\partial A}{\partial \lambda}\frac{\partial B}{\partial
p_{\lambda}}
-\frac{\partial A}{\partial p_{\lambda}}\frac{\partial B}{\partial
\lambda}~~,
\label{11a}
\end{equation}
so as to obtain the canonical 
\begin{equation}
\{x,p_x\}=\{q,p_q\}=
\{\lambda,p_{\lambda}\}=1~~.
\label{12}
\end{equation}
Similarly, given the definition of Eq. (\ref{11a}), we find that the
Poisson brackets of the constraint functions with the Hamiltonian
$H_1$ are given by
\begin{eqnarray}
\{\phi_1,H_1\}&&=
\gamma\omega_1^2\omega_2^2
q-u_2+\phi_1\{\phi_1,u_1\}+\phi_2\{\phi_1,u_2\}~~,
\nonumber \\
\{\phi_2,H_1\}&&=-x+u_1+\phi_1\{\phi_2,u_1\}+\phi_2\{\phi_2,u_2\}~~.
\label{13}
\end{eqnarray}
Consequently, both of the $\{\phi_1,H_1\}$ and $\{\phi_2,H_1\}$ Poisson
brackets will vanish weakly (in the sense of Dirac) if we set
\begin{equation}
u_1=x~~,~~
u_2=\gamma \omega_1^2\omega_2^2q~~.
\label{14}
\end{equation}
On thus imposing these two conditions, $H_1$ is then replaced by a 
new Hamiltonian
\begin{equation}
H_2=\frac{p_x^2}{2\gamma}+\frac{\gamma}{2}(\omega_1^2+\omega_2^2)x^2
-\frac{\gamma}{2}\omega_1^2\omega_2^2q^2+p_qx
+\gamma\omega_1^2\omega_2^2qp_{\lambda}~~.
\label{15}
\end{equation}
With respect to this new Hamiltonian the constraint functions obey
Poisson bracket relations of the form 
\begin{equation}
\{\phi_1,H_2\}=\gamma\omega_1^2\omega_2^2p_{\lambda}~~,~~
\{\phi_2,H_2\}=\{p_{\lambda},H_2\}=0~~.
\label{15a}
\end{equation}
Hence, finally, if we now set
$p_{\lambda}=0$ (so as to enforce $\{\phi_1,H_2\}=0$), the resulting
algebra associated with the four-dimensional $q,p_q,x,p_x$ sector of
the theory will then (as befits a fourth order theory) be closed under
commutation using Poisson brackets defined via 
\begin{equation}
\{A,B\}=\frac{\partial A}{\partial x}\frac{\partial B}{\partial p_x}
-\frac{\partial A}{\partial p_x}\frac{\partial B}{\partial x}
+\frac{\partial A}{\partial q}\frac{\partial B}{\partial p_q}
-\frac{\partial A}{\partial p_q}\frac{\partial B}{\partial q}~~,
\label{15b}
\end{equation}
with the requisite classical Hamiltonian which we
seek then being given by
\begin{equation}
H=\frac{p_x^2}{2\gamma}+p_qx+\frac{\gamma}{2}(\omega_1^2+\omega_2^2)x^2
-\frac{\gamma}{2}\omega_1^2\omega_2^2q^2~~,
\label{16}
\end{equation}
and with the requisite Poisson bracket relations
which define the classical theory being found to be given by
\begin{eqnarray}
\{x,p_x\}&&=1~~,~~\{q,p_q\}=1~~,
\nonumber \\
\{x,H\}=\frac{p_x}{\gamma}~~,~~\{q,H\}=x~~,~~\{p_x,H\}&&
=-p_q-\gamma(\omega_1^2+\omega_2^2)x~~,
~~\{p_q,H\}=\gamma\omega_1^2\omega_2^2q~~.~~~~~~
\label{17}
\end{eqnarray}
With such Poisson bracket relations the canonical equations of motion
thus take the form   
\begin{equation}
\dot{x}=\frac{p_x}{\gamma}~~,~~\dot{q}=x~~,~~\dot{p}_x
=-p_q-\gamma(\omega_1^2
+\omega_2^2)x~~,~~\dot{p}_q=\gamma\omega_1^2\omega_2^2q~~,
\label{18}
\end{equation}
to then enable us to both recover Eq. (\ref{2}) and make
the identification $x=\dot{q}$ in the solution. Additionally,  in
solutions which obey these equations of motion the Legendre
transform $L=p_x\dot{x}+p_q\dot{q}-H$ is found to reduce to Eq.
(\ref{5}) just as it should. The Hamiltonian $H$ of Eq. (\ref{16}) is
thus the correct one for the fourth order theory, and it thus is the one
which is to be quantized.

\section{Connection with the Ostrogradski Hamiltonian }

In considering the theory associated with Eq. (\ref{16}), it
is important to distinguish between the general Hamiltonian $H$ as
defined by Eq. (\ref{16}) and the particular value $H_{\rm STAT}$ that
it takes in the stationary path in which the equations of motion of Eq.
(\ref{18}) are imposed, with the great virtue of the Dirac procedure
being that it allows us to define a classical Hamiltonian $H$ which
takes a meaning (as the canonical generator used in Eq. (\ref{17}))
even for non-stationary field configurations (i.e. even for
configurations for which $p_x\dot{x}+p_q\dot{q}-H$ does not reduce to
the Lagrangian of Eq. (\ref{5})). Thus, for instance, it is the
Hamiltonian of Eq. (\ref{16}) which defines the appropriate phase space
for the problem, so that the path integral for the theory is then
uniquely given by $\int [dq][dp_q][dx][dp_x] {\rm exp}[i\int dt
(p_x\dot{x}+p_q\dot{q}-H)]$ as integrated over a complete set of
classical paths associated with these four independent coordinates and
momenta. 

As regards the stationary
value that the classical $H_{\rm STAT}$ takes when the equations of
motion are imposed, viz.
\begin{equation}
H_{\rm
STAT}=\frac{\gamma}{2}\ddot{q}^2-\frac{\gamma}{2}(\omega_1^2
+\omega_2^2)\dot{q}^2 -\frac{\gamma}{2}\omega_1^2\omega_2^2 q^2-\gamma
\dot{q}
\frac{d^3q}{dt^3}~~,
\label{19}
  \end{equation}
we note that not only is this particular $H_{\rm STAT}$ time
independent (see e.g. Eqs. (\ref{19c}) and (\ref{19d}) below), it is
also recognized as being the Ostrogradski
\cite{Ostrogradski1850} generalized higher derivative Hamiltonian
associated with the Lagrangian of Eq. (\ref{5}), viz. the Hamiltonian
\begin{equation}
H_{\rm OST}=\dot{q}\frac{\partial L}{\partial
\dot{q}}+\ddot{q}\frac{\partial L}{\partial
\ddot{q}}-\dot{q}\frac{d}{dt}\left(\frac{\partial L}{\partial
\ddot{q}}\right) -L 
\label{19a}
\end{equation}
which
Ostrogradski showed to be time independent in solutions which obey
the generalized Euler-Lagrange equation 
\begin{equation}
\frac{\partial
L}{\partial q}-\frac{d}{dt}\left(\frac{\partial L}{\partial
\dot{q}}\right)+\frac{d^2}{dt^2}\left(\frac{\partial L}{\partial
\ddot{q}}\right)=0~~, 
\label{19b}
\end{equation}
while also being its associated time
translation generator \cite{Bak1994}.
Moreover, not only is the classical $H_{\rm STAT}$ time
independent in the unequal frequency solution
$q(t)=a_1e^{-i\omega_1t}+a_2e^{-i\omega_2t}+{\rm c.c.}$ where it
evaluates to 
\begin{equation}
H_{\rm
STAT}(\omega_1\neq\omega_2)=2\gamma(\omega_1^2-\omega_2^2)
(a_1^*a_1\omega_1^2-a_2^*a_2\omega_2^2)~~, 
\label{19c}
\end{equation}
it is even time independent in the equal frequency
$\omega_1=\omega_2=\omega$ temporal runaway solution
$q(t)=c_1e^{-i\omega t}+c_2te^{-i\omega t}+{\rm c.c.}$ where it
evaluates to 
\begin{equation}
H_{\rm
STAT}(\omega_1=\omega_2)=4\gamma
\omega^2\left(2c_2^*c_2+i\omega c_1^*c_2-i\omega c_2^*c_1\right)~~, 
\label{19d}
\end{equation}
with a runaway in time not leading to a runaway in energy. (I.e. there
is an appropriately defined energy which is time independent in
the temporal runaway solution.) 

The value we have obtained for $H_{\rm STAT}(\omega_1=\omega_2)$ in Eq.
(\ref{19d}) recalls to mind the zero energy theorem derived by Boulware,
Horowitz and Strominger \cite{Boulware1983} for fourth order conformal
gravity, another typical fourth order theory. Specifically, Boulware,
Horowitz and Strominger constructed a classical energy for the fourth
order gravity theory, and found that it would vanish identically if the
gravitational field solutions were required to be asymptotically flat. The
analog condition for the equal frequency Pais-Uhlenbeck theory (a prototype
of a pure fourth order field theory) would be to demand the absence of
runaway solutions in time, and thus to set $c_2=0$, a condition which would
then yield for $H_{STAT}(\omega_1=\omega_2)$ none other than the value 
zero. As we thus see, it is the restriction to an analog of asymptotic
flatness which leads to a zero value for the energy in the equal frequency
Pais-Uhlenbeck case, with there being no $c_1^*c_1$ type term present in
$H_{STAT}(\omega_1=\omega_2)$ even though $H_{STAT}(\omega_1\neq
\omega_2)$ contains both $a_1^*a_1$ and $a_2^*a_2$ type terms. Since the
$c_1$ mode can only make a contribution to $H_{STAT}(\omega_1=\omega_2)$
when $c_2$ is non-zero, we can anticipate that in the quantization of the
theory to be presented below the $c_1$ mode will not give rise to a
propagating on-shell state, with the eigenspectrum of the equal frequency
quantum Hamiltonian possessing not the number of energy eigenstates states
associated with a two-dimensional harmonic oscillator (viz. the dipole
structure exhibited in Eq. (\ref{4})), but rather possessing only the
number of energy eigenstates associated with a one-dimensional one. A
further interesting feature of the form we have obtained for $H_{\rm
STAT}(\omega_1=\omega_2)$ in Eq. (\ref{19d}) is that there is choice of
sign for the parameter $\gamma$ for which the pure $c_2$ contribution to the
energy, viz. $8\gamma\omega^2 c_2^*c_2$, is then positive definite. On its
own then, the runaway mode does not give rise to any disease such as a
negative energy, with it yielding a positive one instead. There is thus no
need to require the absence of runaway solutions in the equal frequency
Pais-Uhlenbeck theory (by demanding an analog of asymptotic flatness
\cite{footnote2a}) since in and of themselves they give rise to perfectly
acceptable energies, and do not thus need to be avoided. Since the runaway
solution classical energies are well-behaved, we can thus anticipate that
the quantum energy eigenspectrum associated with the $c_2$ mode sector will
not possess any energy eigenstates with either negative energy or negative
norm, even while the structure found for the unequal frequency
$H_{STAT}(\omega_1\neq \omega_2)$ in Eq. (\ref{19c}) indicates that there
will be energy eigenstates of either negative energy or negative norm in the
quantization of the unequal frequency case.

\section{Constructing the unequal frequency Fock space}

With the Hamiltonian of Eq. (\ref{16}) being defined for both
stationary and non-stationary classical paths, a canonical quantization
of the theory can readily be obtained by replacing ($i$ times) the
Poisson brackets of Eq. (\ref{17}) by canonical equal time commutators.
However, without reference to the explicit structure of the Hamiltonian
itself, we note first that the identification (as suggested but not
required by the equations of motion \cite{footnote3})
\begin{eqnarray}
q(t)&&=a_1e^{-i\omega_1t}+a_2e^{-i\omega_2t}+{\rm H.c.}~~,~~
p_q(t)=i\gamma \omega_1\omega_2^2a_1e^{-i\omega_1t}+
i\gamma \omega_1^2\omega_2 a_2e^{-i\omega_2t}+{\rm H.c.}~~,
\nonumber \\
x(t)&&=-i\omega_1a_1e^{-i\omega_1t}-i\omega_2 a_2e^{-i\omega_2t}+{\rm
H.c.}~~,~~
p_x(t)=-\gamma\omega_1^2a_1e^{-i\omega_1t}-
\gamma \omega_2^2a_2e^{-i\omega_2t}+{\rm H.c.}~~
\nonumber \\
\label{20}
\end{eqnarray}
then furnishes us with a Fock space representation of the
quantum-mechanical commutation relations
\begin{equation}
[x,p_x]=[q,p_q]=i~~,~~[x,q]=[x,p_q]=[q,p_x]=[p_x,p_q]=0
\label{21}
\end{equation}
at all times provided that
\begin{equation}
[a_1,a_1^{\dagger}]=\frac{1}{2\gamma\omega_1
(\omega_1^2-\omega_2^2)}~~,~~
[a_2,a_2^{\dagger}]=\frac{1}{2\gamma\omega_2
(\omega_2^2-\omega_1^2)]}~~,~~
[a_1,a_2^{\dagger}]=0~~,~~[a_1,a_2]=0~~.~~~~
\label{22}
\end{equation}
(Here and throughout both $\omega_1$ and $\omega_2$ are taken to be
positive.)
Then in this convenient Fock representation  the quantum-mechanical
Hamiltonian is found to take the form
\begin{equation}
H=2\gamma(\omega_1^2-\omega_2^2)(\omega_1^2a_1^{\dagger}a_1-
\omega_2^2a_2^{\dagger}a_2)
+\frac{1}{2}(\omega_1+\omega_2)
\label{23}
\end{equation}
with its associated commutators as inferred from Eq. (\ref{17}) then
automatically being satisfied. With the quantity
$\gamma(\omega_1^2-\omega_2^2)$  being taken to be positive for
definitiveness, we see that the $[a_2,a_2^{\dagger}]$ commutator is
negative definite, and with $H$ being diagonal in the $a_1,a_2$
occupation number basis, we see that the state defined  by
\begin{equation}
a_1 |\Omega\rangle=0~~,~~a_2|\Omega\rangle =0
\label{24}
\end{equation}
is its ground state,\cite{footnote4} that the states
\begin{equation}
|+1\rangle =[2\gamma\omega_1(\omega_1^2-\omega_2^2)]^{1/2}
a_1^{\dagger}|\Omega\rangle~~,~~
|-1\rangle =[2\gamma\omega_2(\omega_1^2-\omega_2^2)]^{1/2}
a_2^{\dagger}|\Omega\rangle
\label{25}
\end{equation}
are both positive energy eigenstates with respective energies $\omega_1$
and $\omega_2$ above the ground state, that the state $|+1\rangle$ has a
norm equal to plus one, but that the state $|-1\rangle$ has norm minus
one, a ghost state. Thus, as anticipated, the correct Hamiltonian for
the unequal frequency theory can be diagonalized in a basis of positive
and negative norm states, with the relevant negative norm state wave
functions being explicitly constructed via Eq. (\ref{25}) and its
multi-particle generalizations. With the eigenstates of $H$ labeling the
asymptotic states associated with scattering in the presence of any
interaction Lagrangian $L_I$ which might be added on to the original
Lagrangian $L$ of the theory, the effect of $L_I$ would be expected to
induce transitions between asymptotic in and out states of opposite
norm, with the unequal frequency theory then being
non-unitary.\cite{footnote5}  However, even though the Hamiltonian of
the unequal frequency theory does have ghost eigenstates, since the
commutation relations given in Eq. (\ref{22}) become singular in the
equal frequency limit while both the Hamiltonian of Eq. (\ref{23}) and
the normalized $|+1\rangle$ and
$|-1\rangle$ states develop zeroes, we will have to exercise some
caution in trying to discover exactly what happens to the ghost states
when this limit is taken.

\section{Constructing the equal frequency Fock space}

To explore the equal frequency limit it is convenient to
introduce new Fock space variables according to
\begin{eqnarray}
&&a_1=\frac{1}{2}\left(a-b+\frac{2b\omega}{\epsilon}\right)~~,~~
a_2=\frac{1}{2}\left(a-b-\frac{2b\omega}{\epsilon}\right)~~,
\nonumber \\
&&a=a_1\left(1+\frac{\epsilon}{2\omega}\right)
+a_2\left(1-\frac{\epsilon}{2\omega}\right)~~,~~b=
\frac{\epsilon}{2\omega}(a_1-a_2)~~.
\label{26a}
\end{eqnarray}
where
\begin{equation}
\omega=\frac{(\omega_1+\omega_2)}{2}~~,~~\epsilon=\frac{(\omega_1
-\omega_2)}{2}
\label{26b}
\end{equation}
These variables are found to obey commutation relations of the form
\begin{equation}
[a,a^{\dagger}]=\lambda~~,~~
[a,b^{\dagger}]=\mu~~,~~[b,a^{\dagger}]=\mu~~,~~
[b,b^{\dagger}]=\nu~~,~~[a,b]=0~~,
\label{26c}
\end{equation}
where
\begin{equation}
\lambda=\nu=-\frac{\epsilon^2}{16\gamma(\omega^2-\epsilon^2)\omega^3}~~,~~
\mu=\frac{(2\omega^2-\epsilon^2)}{
16\gamma(\omega^2-\epsilon^2)\omega^3}~~.
\label{26}
\end{equation}
As such, the introduction of the operators
$a,~a^{\dagger},~b,~b^{\dagger}$ is initially nothing more than a
rewriting of the original $a_1,~a_1^{\dagger},~a_2,~a_2^{\dagger}$
operators. Their utility derives from the fact that when expressed in
terms of them the coordinate $q(t)$ of Eq. (\ref{20}) gets rewritten as
\begin{equation}
q(t)=e^{-i\omega t}\left[(a-b){\rm cos}~\epsilon t-
\frac{2ib\omega}{\epsilon} {\rm sin}~\epsilon t\right] +{\rm H.c.}~~,
\label{27}
\end{equation}
and thus has a well defined $\epsilon \rightarrow 0$ limit, viz.
\begin{equation}
q(t,\epsilon =0)=e^{-i\omega t}(a-b-2ib \omega t) +{\rm H.c.}~~.
\label{28}
\end{equation}
Similarly, the Hamiltonian of Eq. (\ref{23}) which takes the form   
\begin{equation}
H=8\gamma \omega^2\epsilon^2(a^{\dagger}a-b^{\dagger}b)+
8\gamma \omega^4 (2b^{\dagger}b+a^{\dagger}b
+ b^{\dagger}a)+\omega
\label{28a}
\end{equation}
in the new variables, then 
limits to 
\begin{equation}
H(\epsilon=0)=8\gamma \omega^4 (2b^{\dagger}b+a^{\dagger}b
+ b^{\dagger}a)+\omega~~,
\label{29}
\end{equation}
while the following commutators of interest have limiting 
form 
\begin{eqnarray}
&&[H(\epsilon=0),a^{\dagger}]
=\omega (a^{\dagger}+2b^{\dagger})~~,~~
[H(\epsilon=0),a]=-\omega (a+2b)~~,
\nonumber \\
&&[H(\epsilon=0),b^{\dagger}]=\omega b^{\dagger}~~,~~
[H(\epsilon=0),b]=-\omega b~~,
\nonumber \\
&&[a+b,a^{\dagger}+b^{\dagger}]=2\hat{\mu}~~,~~
[a-b,a^{\dagger}-b^{\dagger}]=-2\hat{\mu}~~,
~~[a+b,a^{\dagger}-b^{\dagger}]=0~~,
\label{30}
\end{eqnarray}
where now $\mu(\epsilon=0)=\hat{\mu} =1/(8\gamma \omega^3)$.
Thus even while the relation between the $(a,~b)$ and $(a_1,~a_2)$ sets
of operators is explicitly singular in the limit, the limiting
prescription obtained by using the $a$ and $b$ operators of Eq.
(\ref{26a}) nonetheless leads to the $\epsilon=0$ operator algebra
given in Eqs. (\ref{29}) and (\ref{30}) in which there are no singular
terms at all. However, this is not the only way to take the $\epsilon
\rightarrow 0$ limit, as we could instead take the limit not of the
operators of the unequal frequency theory, but rather of the states.
There are thus two possible ways to construct the equal frequency
theory --  we can either start with the $\epsilon=0$ operator algebra
and construct a new Fock space for it from scratch, or we can construct
the equal frequency states as the $\epsilon \rightarrow 0$ limit of the
unequal frequency Fock space states. With these two prescriptions not
being equivalent, we shall explore both of them, and discuss first the
limit of the operator algebra.

\subsection{Taking the limit of the operator algebra}

For the operator algebra limit, we use the $\epsilon \rightarrow 0$
limit of the $a,~a^{\dagger},~b,~b^{\dagger}$ operators as the
dynamical variables, with use of these operators then enabling us to
construct an $\epsilon=0$ theory which is well defined. For this theory
the $\epsilon=0$ Fock space is then built on the Fock vacuum
$|\Omega\rangle$ defined by $a|\Omega\rangle=b|\Omega\rangle=0$, (we use
this basis since, according to Eq. (\ref{22}), states built on the
unequal frequency Fock vacuum in which
$a_1|\Omega\rangle=0,~a_2|\Omega\rangle=0$ become undefined in the
$\epsilon=0$ limit), with the state $|\Omega \rangle$ being an
eigenstate of $H$ with energy $\omega$. Moreover, in this
$\epsilon\rightarrow 0$ limit we see that the $[a,a^{\dagger}]$ and
$[b,b^{\dagger}]$ commutators both vanish, with, as we shall see below,
there actually being two ways rather than one in which the Fock space
states can implement this vanishing depending on how the
$\epsilon\rightarrow 0$ limit is actually taken. However, before
constructing either of these two ways, we note that because the effect
of the Hamiltonian is to shift
$a^{\dagger}$ by $2b^{\dagger}$ in Eq. (\ref{30}), and because (unlike
the unequal frequency case) neither of the operators ($a^{\dagger} \pm
b^{\dagger}$) which diagonalize the Fock space basis acts as a ladder
operator for the Hamiltonian,\cite{footnote5a} we can
anticipate that the eigenspectrum of $H(\epsilon=0)$ will be quite
different from the unequal frequency eigenspectrum.  

Because $b^{\dagger}$ does act as a ladder operator for $H(\epsilon=0)$,
it is possible to construct a one particle energy eigenstate, viz. the
state $b^{\dagger}|\Omega\rangle$ with energy $2\omega$, but if we try
to normalize it (our first way to realize the $\epsilon=0$ Fock space)
the vanishing of the $[b,b^{\dagger}]$ commutator would entail that this
state would have to have zero norm. On its own a zero norm state
(unlike a negative norm state) does not lead to loss of probability,
but its very existence entails the existence of negative norm states
elsewhere in the theory.\cite{footnote6} Thus, with the state
$a^{\dagger}|\Omega\rangle$ also having zero norm, we immediately
construct the states
\begin{equation}
|\pm\rangle = \frac{(a^{\dagger}\pm
b^{\dagger})}{(2\hat{\mu})^{1/2}}|\Omega\rangle~~,
\label{31}
\end{equation}
states which obey
\begin{equation}
\langle +|+\rangle =1~~, ~~\langle -|-\rangle =-1~~, ~~\langle
+|-\rangle =0~~.
\label{32}
\end{equation}
However, unlike the orthogonal positive and negative norm states
$|\pm1\rangle$ of Eq. (\ref{25}) which are eigenstates of the unequal
frequency Hamiltonian, this time we find that neither of the
$|\pm\rangle$ states is an eigenstate of the $\epsilon=0$ Hamiltonian
since
\begin{equation}
H(\epsilon=0)|\pm\rangle=2\omega|\pm\rangle +4\omega
b^{\dagger}|\Omega\rangle~~.
\label{33}
\end{equation}
Thus even while the $\epsilon=0$ Fock space possesses negative norm
ghost states, this time they can only exist off shell (where they can
still regulate Feynman diagrams) but cannot materialize as on shell
asymptotic in and out states. With a similar situation being found in
the two particle sector \cite{footnote7} where only the state
$(b^{\dagger})^2|\Omega\rangle$ with energy $3\omega$ is an energy
eigenstate, we see that the only states which are eigenstates
of $H(\epsilon=0)$ are those of the form
$(b^{\dagger})^n|\Omega\rangle$, with all such states having zero
norm and positive energy.\cite{footnote8}

Now at first it is quite perplexing that the $\epsilon=0$ theory
possesses far fewer energy eigenstates than the $\epsilon \neq 0$
theory, possessing only the number of eigenstates associated with a
one-dimensional harmonic oscillator rather than a two-dimensional
one. The reason for such an outcome derives from the fact that
while the normal situation for square matrices is that the number of
independent eigenvectors of a square matrix is the same as the
dimensionality of the matrix, there are certain matrices, known as
defective matrices, for which this is not in fact the case. A typical
example of such a defective matrix is the non-Hermitian, Jordan block
form, two-dimensional matrix
\begin{eqnarray}
M= \pmatrix{1&1 \cr 0&1}~~.
\label{32a}
\end{eqnarray}
Specifically, while this matrix has two eigenvalues both of which are
real (despite the lack of hermiticity) and equal to $1$ (the trace of
$M$ is equal to $2$ and its determinant is equal to $1$), solving the
equation 
\begin{eqnarray}
\pmatrix{1&1 \cr 0&1}\pmatrix{p \cr q}=\pmatrix{p+q\cr q}=\pmatrix{p
\cr q}
\label{32b}
\end{eqnarray}
leads only to $q=0$, with there thus only being one eigenvector
despite the two-fold degeneracy of the eigenvalue, with the space on
which the matrix $M$ acts not being complete.\cite{footnote8a}

For the case of interest to us here, the action of the
unequal frequency Hamiltonian of Eq. (\ref{28a}) on the one-particle
states $a^{\dagger}|\Omega\rangle$, $b^{\dagger}|\Omega\rangle$ yields
\begin{eqnarray}
Ha^{\dagger}|\Omega\rangle=&&
\frac{1}{2\omega}\left[(4\omega^2+\epsilon^2)a^{\dagger} |\Omega\rangle 
+(4\omega^2-\epsilon^2)b^{\dagger}|\Omega\rangle\right]
\nonumber \\
Hb^{\dagger}|\Omega\rangle
=&&\frac{1}{2\omega}\left[\epsilon^2a^{\dagger} |\Omega\rangle 
+(4\omega^2-\epsilon^2)b^{\dagger}|\Omega\rangle\right]~~.
\label{32c}
\end{eqnarray}
For this sector we can define a matrix
\begin{eqnarray}
M(\epsilon)=
\frac{1}{2\omega}\pmatrix{4\omega^2+\epsilon^2&4\omega^2-\epsilon^2
\cr \epsilon^2&4\omega^2-\epsilon^2}
\label{32d}
\end{eqnarray}
whose eigenvalues are given as $2\omega+\epsilon$ and
$2\omega -\epsilon$ (the trace of $M(\epsilon)$ is $4\omega$ and its
determinant is $4\omega^2-\epsilon^2$). For such eigenvalues, 
energy eigenvectors which obey  
\begin{eqnarray}
H|2\omega \pm \epsilon\rangle =(2\omega \pm \epsilon)|2\omega \pm
\epsilon\rangle 
\label{32e}
\end{eqnarray}
are then readily constructed as
\begin{equation}
|2\omega \pm \epsilon\rangle =\left[\pm\epsilon a^{\dagger}+(2\omega
\mp \epsilon)b^{\dagger}\right]|\Omega \rangle~~.
\label{32f}
\end{equation}
As we see, as long as $\epsilon \neq 0$, the two one-particle sector 
eigenvectors of the Hamiltonian $H$ are distinct. However when we let
$\epsilon$ go to zero, the two eigenvectors in Eq. (\ref{32f}) collapse
onto a single eigenvector, viz. the vector $b^{\dagger}|\Omega \rangle$,
the two eigenvalues collapse onto a common eigenvalue, viz. $2\omega$,
and the matrix $M(\epsilon=0)$ of Eq. (\ref{32d}) becomes
($2\omega$ times) the defective one given in Eq. (\ref{32a}).
Consequently, while the dimensionality of the
$a,~a^{\dagger},~b,~b^{\dagger}$ based Fock space does not change in the
$\epsilon \rightarrow 0$ limit (i.e. the dimensionality of the Fock space
remains that of a two-dimensional harmonic oscillator), nonetheless, in
the limit the Hamiltonian becomes defective, with $H(\epsilon=0)$ thus
possessing far fewer eigenvectors than $H(\epsilon\neq 0)$, and with the
dimensionality of the space of energy eigenvectors of $H(\epsilon=0)$
actually being the same as that of a one-dimensional rather than a
two-dimensional harmonic oscillator.\cite{footnote8b} As we thus see, the
$\epsilon
\rightarrow 0$ limit is indeed highly singular. Finally, with Eq.
(\ref{32c}) reducing to 
\begin{eqnarray}
H(\epsilon=0)a^{\dagger}|\Omega\rangle=&&
2\omega\left[a^{\dagger} |\Omega\rangle 
+b^{\dagger}|\Omega\rangle\right]
\nonumber \\
H(\epsilon=0)b^{\dagger}|\Omega\rangle
=&&2\omega b^{\dagger}|\Omega\rangle
\label{32g}
\end{eqnarray}
when $\epsilon=0$, we confirm again that $H(\epsilon=0)$ only has one
eigenvector, the zero norm $b^{\dagger}|\Omega\rangle$ with eigenvalue
$2\omega$.

As well as these zero norm states, the $\epsilon=0$ Fock space also
possesses harmonic oscillator coherent states as well. To construct
these particular states it is convenient to introduce new operators
\begin{equation}
\alpha=a+b~~,~~\beta=a-b
\label{33a}
\end{equation}
which obey 
\begin{equation}
[\alpha,\alpha^{\dagger}]=2\hat{\mu}~~,~~[\beta,\beta^{\dagger}]
=-2\hat{\mu}~~,~~ [\alpha,\beta^{\dagger}]=0~~,~~
[\beta,\alpha^{\dagger}]=0~~,~~[\alpha,\beta]=0~~.~~~
\label{33b}
\end{equation}
In terms of these operators the Hamiltonian of Eq. (\ref{29}) may be
rewritten as
\begin{equation}
H(\epsilon=0)=\frac{\omega}{2\hat{\mu}}\left[2\alpha^{\dagger}\alpha
-\alpha^{\dagger}\beta-\beta^{\dagger}\alpha\right] +\omega~~,
\label{33c}
\end{equation}
with $\alpha^{\dagger}$ and $\beta^{\dagger}$ then being
found to obey the relations
\begin{eqnarray}
&&[H(\epsilon=0),\alpha^{\dagger}]
=\omega(2\alpha^{\dagger}-\beta^{\dagger})~~,~~
[H(\epsilon=0),\beta^{\dagger}]=\omega\alpha^{\dagger}~~,
\nonumber \\
&&[H(\epsilon=0),\alpha^{\dagger}-\beta^{\dagger}]
=\omega(\alpha^{\dagger}-\beta^{\dagger})~~,~~
[H(\epsilon=0),\alpha-\beta]=-\omega(\alpha-\beta)~~.
\label{33d}
\end{eqnarray}
In terms of these operators we construct the coherent state
\begin{equation}
|g\rangle=e^{\alpha^{\dagger}\beta^{\dagger}}|\Omega\rangle
\label{33e}
\end{equation}
where the vacuum is again the one which $a$ and $b$, and thus $\alpha$
and $\beta$, annihilate. Using the well-known relation
$e^{A+B}=e^{A}e^{B}e^{-[A,B]/2}=e^{B}e^{A}e^{-[B,A]/2}$ which holds
when $[A,[A,B]]=[B,[A,B]]=0$, it can readily be shown that $|g\rangle$
has positive norm, viz. $\langle g|g\rangle=\langle \Omega|
e^{\alpha^{\dagger}\beta^{\dagger}}e^{\alpha\beta}
e^{(\beta^{\dagger}\beta-\alpha^{\dagger}
\alpha -1)}|\Omega\rangle =1/e$.
Similarly, using the well-known relation
$e^{A}Be^{-A}=B+[A,B]+[A,[A,B]]/2 +[A,[A,[A,B]]]/6+...$, it can
also be shown that application of the Hamiltonian to this state
yields
\begin{equation}
H(\epsilon=0)|g\rangle=\omega(\alpha^{\dagger 2}
-\beta^{\dagger 2}+2\alpha^{\dagger}\beta^{\dagger})
|g\rangle~~,
\label{33f}
\end{equation}
with the coherent state thus not being an energy eigenstate. This
same analysis generalizes to any other state of the form of an
exponential of a string of creation operators acting on the vacuum,
since the commutator of $H(\epsilon=0)$ with the string yields another
string consisting also purely of creation operators, a string which
thus commutes with the original string of creation operators. Thus
whether we use states with a definite number of particles (such
$a^{\dagger m}b^{\dagger n}|\Omega \rangle$) or states with an
indefinite number of particles (such as $|g \rangle$), we find that
only the $b^{\dagger n}|\Omega \rangle$ states are energy
eigenstates, with all the $\epsilon \neq 0$ negative norm states
having moved off shell in the limit. Consequently, in the $\epsilon=0$
theory there are no energy eigenstates with negative norm or negative
energy. 

As well as this, another interesting aspect of the
$\epsilon=0$ theory it is that even while it is a fourth order theory,
the $\epsilon=0$ theory only possesses the number of observable states
that would be found in an ordinary second order theory. Consequently,
in a field-theoretic generalization of the equal frequency
Pais-Uhlenbeck fourth order oscillator, we would anticipate that there
would only be one observable particle in the theory rather than the two
that would be associated with the field-theoretic analog of the unequal
frequency case as exhibited schematically in Eq. (\ref{4}).  Replacing a
pure second order field theory by a pure fourth order one would thus
not be expected to bring about any change in the number of observable
states.

Since the above equal frequency theory possesses zero norm states, as a
quantum-mechanical theory it is thus somewhat unconventional. Even though
the physical viability of the theory is not in question since the theory
has been shown to possess no energy eigenstates with negative norm,
nonetheless, it is still of interest to try to recast the theory in an
alternate form in which the theory has a more conventional look to it.
Thus we seek to reinterpret the zero norm states as positive norm ones by
using an unconventional scalar product with which to define the norm. To
this end, we note that in constructing the eigenstates of a Hamiltonian,
all that is needed is the introduction of a Hilbert space on which the
Hamiltonian is to act to the right on a set of ket vectors $|\psi
\rangle$, with there being no need to introduce the dual vector bra
states of those ket vectors for this purpose, with the structure of the
energy eigenspectrum thus not being in any way sensitive to the structure
of the dual space vectors.\cite{footnote8c} The choice
of definition of a scalar product is thus independent of the structure
of the energy eigenspectrum, and while one ordinarily defines the dual bra
vector $\langle \psi|$ simply as the conjugate of $|\psi \rangle$, other
choices are possible, and all of them provide legitimate formulations of
quantum mechanics.\cite{footnote8d}  

Thus for our purposes here, we note that with both of the equal
frequency theory $a^{\dagger}|\Omega\rangle$ and $b^{\dagger}|\Omega
\rangle$ states being zero norm states if their respective conjugates are
canonically defined as $\langle \Omega| a$, $\langle \Omega| b$, in
terms of the operators $\alpha$ and $\beta$ which diagonalize the Fock
algebra, we instead define the dual of the state $\beta^{\dagger}|\Omega
\rangle$ to not be the standard $\langle \Omega| \beta$ but to be minus
one times it (viz. $-\langle \Omega|\beta$) instead, while at the same
time we continue to keep the dual of $\alpha^{\dagger}|\Omega \rangle$ to
be the standard $\langle \Omega|\alpha$.  With such a dual state
definition, the dual of 
$b^{\dagger}|\Omega\rangle=(1/2)(\alpha^{\dagger}-\beta^{\dagger})|\Omega
\rangle$ is given by $(1/2)\langle \Omega|(\alpha+\beta)$, viz. by
$\langle \Omega|a$, with the overlap of 
$b^{\dagger}|\Omega\rangle$ with its own dual
then being given by $(1/4)\langle \Omega|
(\alpha+\beta)(\alpha^{\dagger}-\beta^{\dagger})|\Omega
\rangle=\hat{\mu}=1/8\gamma\omega^3$, a norm which, for positive
$\gamma$, is then positive rather than zero. And similarly, the dual of 
$a^{\dagger}|\Omega\rangle=(1/2)(\alpha^{\dagger}+\beta^{\dagger})|\Omega
\rangle$ is given by $(1/2)\langle \Omega|(\alpha-\beta)$, viz. by
$\langle \Omega|b$, with the overlap
of $a^{\dagger}|\Omega \rangle$ with its own dual
then being given by $(1/4)\langle \Omega|
(\alpha-\beta)(\alpha^{\dagger}+\beta^{\dagger})|\Omega
\rangle=\hat{\mu}=1/8\gamma\omega^3$, a norm which is then also positive
for the same choice of sign of $\gamma$. Moreover, the overlap of the
dual of $b^{\dagger}|\Omega \rangle$ with $a^{\dagger}|\Omega \rangle$ is
given by 
$(1/4)\langle \Omega|
(\alpha+\beta)(\alpha^{\dagger}+\beta^{\dagger})|\Omega
\rangle$, an overlap which is zero,
while the overlap of the
dual of $a^{\dagger}|\Omega \rangle$ with $b^{\dagger}|\Omega
\rangle$ is given by 
$(1/4)\langle \Omega|
(\alpha-\beta)(\alpha^{\dagger}-\beta^{\dagger})|\Omega
\rangle$, an overlap which is zero also. With respect to this
definition of scalar product then, the states $a^{\dagger}|\Omega
\rangle$ and $b^{\dagger}|\Omega \rangle$ form an orthonormal basis in
which all states have positive norm. Thus because we do not
need to specify the dual vectors in order to construct energy eigenkets,
we have flexibility in defining what we mean by a
norm and a scalar product, and can thus make the equal frequency theory
zero norm states become orthonormal positive norm states instead,
and can do so without in any way altering the fact that
$b^{\dagger}|\Omega\rangle$ is an energy eigenket while
$a^{\dagger}|\Omega \rangle$ is not. 

In addition to the above formal construction of dual vectors, there is
also an explicit operational way to implement it. Specifically, we note
that before we seek to define a new scalar product, the standard scalar
products are given by 
$\langle \Omega|aa^{\dagger}|\Omega\rangle=0$, 
$\langle \Omega|ab^{\dagger}|\Omega\rangle=\hat{\mu}$,
$\langle \Omega|ba^{\dagger}|\Omega\rangle=\hat{\mu}$,
$\langle \Omega|bb^{\dagger}|\Omega\rangle=0$. We now introduce an
operator $C$ whose action on $\alpha^{\dagger}|\Omega\rangle$ and
$\beta^{\dagger}|\Omega\rangle$ is of the form
$C\alpha^{\dagger}|\Omega\rangle=\alpha^{\dagger}|\Omega\rangle$,
$C\beta^{\dagger}|\Omega\rangle=-\beta^{\dagger}|\Omega\rangle$. In
consequence, its action on $a^{\dagger}|\Omega\rangle$ and 
$b^{\dagger}|\Omega\rangle$ is of the form
$Ca^{\dagger}|\Omega\rangle=b^{\dagger}|\Omega\rangle$,
$Cb^{\dagger}|\Omega\rangle=a^{\dagger}|\Omega\rangle$, an action which
thus interchanges $a^{\dagger}|\Omega\rangle$ and 
$b^{\dagger}|\Omega\rangle$ with each other. With respect to this operator
$C$ we thus define scalar products which evaluate to
$\langle \Omega|aCa^{\dagger}|\Omega\rangle=\hat{\mu}$, 
$\langle \Omega|aCb^{\dagger}|\Omega\rangle=0$,
$\langle \Omega|bCa^{\dagger}|\Omega\rangle=0$,
$\langle \Omega|bCb^{\dagger}|\Omega\rangle=\hat{\mu}$. Thus with
$\langle \Omega|aC$ being defined as the dual of
$a^{\dagger}|\Omega\rangle$, and with
$\langle \Omega|bC$ being defined as the dual of
$b^{\dagger}|\Omega\rangle$, we thus generate an orthonormal basis. Now
we saw above that the dual of 
$a^{\dagger}|\Omega\rangle$ is given as $\langle \Omega|b$ while 
the dual of 
$b^{\dagger}|\Omega\rangle$ is given as $\langle \Omega|a$.
In this construction then we thus identify the dual of
$a^{\dagger}|\Omega\rangle$ as 
$\langle \Omega|aC=\langle \Omega|b$, and the dual of
$b^{\dagger}|\Omega\rangle$ as 
$\langle \Omega|bC=\langle \Omega|a$. Since the role of $C$ is to
interchange $a^{\dagger}|\Omega\rangle$ and $b^{\dagger}|\Omega\rangle$,
we can thus conveniently describe the $a^{\dagger}|\Omega\rangle$, 
$b^{\dagger}|\Omega\rangle$ 2-space in the language of Pauli spinors.
Thus on defining $a^{\dagger}|\Omega\rangle=|\uparrow
\rangle$, $b^{\dagger}|\Omega\rangle=|\downarrow \rangle$, we see
that the action of $C$ on $a^{\dagger}|\Omega\rangle$ and 
$b^{\dagger}|\Omega\rangle$ can be written as
$C|\uparrow\rangle=|\downarrow \rangle$,
$C|\downarrow\rangle=|\uparrow \rangle$. Similarly, its action on
$\alpha^{\dagger}|\Omega\rangle=(|\uparrow\rangle+|\downarrow \rangle)$
and 
$\beta^{\dagger}|\Omega\rangle=(|\uparrow\rangle-|\downarrow \rangle)$ can
be written as
$C(|\uparrow\rangle+|\downarrow \rangle)=(|\uparrow\rangle+|\downarrow
\rangle)$,
$C(|\uparrow\rangle-|\downarrow \rangle)=-(|\uparrow\rangle-|\downarrow
\rangle)$. In the $a^{\dagger}|\Omega\rangle$, 
$b^{\dagger}|\Omega\rangle$ 2-space then the operator $C$ can be
represented by none other than the Pauli matrix $\sigma_1$, to thus
establish that the needed $C$ does indeed exist. Then given this
definition of conjugate states, if we define wave functions via
$\psi_a(x)=\langle x|a^{\dagger}|\Omega \rangle$, $\psi_b(x)=\langle
x|b^{\dagger}|\Omega \rangle$ with respective conjugates 
$\psi_a^*(x)=\langle \Omega |aC|x\rangle$, $\psi_b^*(x)=\langle \Omega
|bC|x\rangle$, we are able to obtain the conventional orthonormality
relations for wave functions, viz. $\int dx\psi_a^*(x)\psi_a(x)=\int dx
\langle\Omega |aC|x\rangle\langle x|a^{\dagger}|\Omega \rangle=
\langle \Omega |aCa^{\dagger}|\Omega \rangle=\hat{\mu}$, 
$\int dx\psi_a^*(x)\psi_b(x)=0$,
$\int dx\psi_b^*(x)\psi_a(x)=0$,
$\int dx\psi_b^*(x)\psi_b(x)=\hat{\mu}$.

As regards the Hamiltonian of the theory, we note that with respect to
our definition of conjugate states, its matrix elements evaluate to  
$\langle \Omega|aCHa^{\dagger}|\Omega\rangle=1/4\gamma\omega^2$,
$\langle \Omega|aCHb^{\dagger}|\Omega\rangle=0$,
$\langle \Omega|bCHa^{\dagger}|\Omega\rangle=1/4\gamma\omega^2$,
$\langle \Omega|bCHb^{\dagger}|\Omega\rangle=1/4\gamma\omega^2$. As we
see, with respect to this definition of conjugates, the Hamiltonian
remains defective, with its diagonal elements still being real. With
there thus still only being one energy eigenstate, viz. the original
$b^{\dagger}|\Omega\rangle$, with which to characterize asymptotic in and
out S-matrix states, the theory remains unitary not because $H$ has
became Hermitian with respect to the new definition of conjugate states,
but because it has remained defective. While we have thus shown that one
could equivalently either work with an equal frequency theory with
zero norm states or recast the theory as one with positive norm states
instead, to conclude this paper we now present an entirely different
approach to the equal frequency theory, one that will lead to positive
norm states without any need to introduce non-canonical dual vector
states at all.

\subsection{Taking the limit of the Hilbert space} 

While we have seen that it is possible to define an unconventional scalar
product with respect to which all of the $\epsilon=0$ theory zero norm
Fock states then have positive norm instead, it is also of interest to see
if we could construct an alternate $\epsilon \rightarrow 0$ limiting
theory in which all energy eigenstates would have a positive norm even
when the conventional definition of conjugate states is used. To this end
we note that since we obtained zero norm states in the above by first
going to the $\epsilon=0$ limit of the operator algebra before
constructing the states, to avoid such zero norm states an alternate
procedure would be to first construct a set of states with non-zero norm,
and then take the $\epsilon \rightarrow 0$ limit while holding the norms
of these states fixed. With the algebra of the $a$ and $b$ operators
already being defined in Eq. (\ref{26c}) even prior to taking the
$\epsilon \rightarrow 0$ limit, we can thus construct a basis for the
$\epsilon \neq 0$ Fock space via states built on a vacuum which obeys
$a|\Omega\rangle=b|\Omega\rangle=0$ rather than on one which obeys
$a_1|\Omega\rangle=0,~a_2|\Omega\rangle=0$, and then explore the 
$\epsilon \rightarrow 0$ limit of those particular states. We thus
build normalized $\epsilon\neq 0$ states such as (the choice $\gamma<0$
assures the positivity of $\lambda$ and $\nu$) the one particle
\begin{equation}
|1,0\rangle=\frac{a^{\dagger}}{\lambda^{1/2}}|\Omega\rangle~~,~~
|0,1\rangle=\frac{b^{\dagger}}{\nu^{1/2}}|\Omega\rangle~~,
\label{34}
\end{equation}
the two particle
\begin{equation}
|2,0\rangle=\frac{(a^{\dagger})^2}{2^{1/2}\lambda}|\Omega\rangle~~,~~
|1,1\rangle=\frac{a^{\dagger}b^{\dagger}}{(\mu^2
+\lambda\nu)^{1/2}}|\Omega\rangle~~,~~
|0,2\rangle=\frac{(b^{\dagger})^2}{2^{1/2}\nu}|\Omega\rangle~~,
\label{35}
\end{equation}
and so on.

Even while this particular basis is not a basis of eigenstates of
the $\epsilon \neq 0$ Hamiltonian, for the $\epsilon \neq 0$ theory
this particular basis is just as complete a basis as the one built
out of the $\epsilon \neq 0$ eigenstates. However, unlike the
$\epsilon \neq 0$ energy eigenstate basis, remarkably, and
crucially for our purposes here as it will turn out, every single state
constructed via the $\epsilon \neq 0$ $a^{\dagger}$ and $b^{\dagger}$
operators is found to have positive norm. It is not however an
orthogonal basis since overlaps such as
$\langle 1,0|0,1\rangle=\mu/(\lambda \nu)^{1/2}$ are non-zero, overlaps
which actually become singular in the $\epsilon \rightarrow 0$ limit.
Since this basis is complete we can use it to define a particular
limiting procedure in which the normalization of each of the particle
states in Eqs. (\ref{34}) and (\ref{35}) and their multi-particle
generalization is held fixed while $\epsilon$ is allowed to go to zero.
Then in such a limit we find from Eq. (\ref{34}) that
$a^{\dagger}|\Omega\rangle$ and $b^{\dagger}|\Omega\rangle$ both
become null vectors. However, despite this, we cannot conclude that
the creation operators annihilate the vacuum identically in this
limit since matrix elements such as
$\langle \Omega|[a,b^{\dagger}]|\Omega\rangle = \mu $ do not vanish
in the limit ($\mu(\epsilon=0)=\hat{\mu}\neq 0$), with  product
operator actions such as $a$ acting on $b^{\dagger}|\Omega\rangle$
being singular. Instead, the creation operators must be thought of as
annihilating the vacuum weakly (i.e. in some but not all matrix
elements), but not strongly as an operator identity. Now since
neither of the states $a^{\dagger}|\Omega\rangle$ or
$b^{\dagger}|\Omega\rangle$ survives in the limit, $H(\epsilon=0)$
now has no one particle eigenstates at all. With the two particle
states $(a^{\dagger})^2|\Omega\rangle$ and
$(b^{\dagger})^2|\Omega\rangle$ also becoming null in the limit, the
only two particle state which is found to survive in this limit is
the positive norm state
$a^{\dagger}b^{\dagger}|\Omega\rangle/\hat{\mu}$ (since the
$[a,b^{\dagger}]$ commutator does not vanish), and, quite remarkably,
in this limit the state is also found to actually become an energy
eigenstate, viz.
\begin{equation}
H(\epsilon=0)\frac{a^{\dagger}b^{\dagger}}{\hat{\mu}}|\Omega\rangle
=\frac{3\omega}{\hat{\mu}} a^{\dagger}b^{\dagger}|\Omega\rangle+
\frac{2\omega}{\hat{\mu}} (b^{\dagger})^2|\Omega\rangle\equiv
\frac{3\omega}{\hat{\mu}} a^{\dagger}b^{\dagger}|\Omega\rangle~~,
\label{36}
\end{equation}
with the energy of this two-particle state being equal to $3\omega$,
which is nicely positive. With this analysis immediately generalizing to
the higher multiparticle states as well, we see that the only states
which then survive in the limit are states of the form
$(a^{\dagger}b^{\dagger})^n|\Omega\rangle$, positive norm states
which also become positive energy eigenstates in the limit, with the
observable sector of the theory thus being completely acceptable.  We
thus recognize two limiting procedures, first taking the limit of the
algebra and then constructing the Fock space, or first constructing
the Fock space and then taking the limit of its states, with this
latter limiting procedure being an extremely delicate one in which the
only states which survive as observable ones are
composite.\cite{footnote9} Thus, to conclude, we see that in the equal
frequency limit the quantum theory based on the fourth order Lagrangian
of Eq. (\ref{5}) is in fact completely acceptable (in fact, technically,
what we have shown is that the presence of energy eigenstates with
negative norm in the unequal frequency theory is simply not a reliable
indicator as to their possible presence in the equal frequency case),
and it would thus be of interest to see the degree to which our results
here might carry over to full fourth order
field theories.\cite{footnote10}

\begin{acknowledgments}
One of the authors (P.D.M) wishes to thank Drs. E. E. Flanagan and
J. Javanainen for useful discussions. The work of P.D.M. has been
supported in part by the Department of Energy under grant No.
DE-FG02-92ER4071400.
\end{acknowledgments}


\begin{thebibliography}{}
\bibitem{Stelle1977} K. S. Stelle, Phys. Rev. D {\bf 16}, 953 (1977);
Gen. Relativ. Gravit. {\bf 9}, 353 (1978).

\bibitem{footnote0} With pure second order theories not
possessing ghosts, and with second plus fourth order theories
possessing them, it might be presumed that the ghosts arise from the
pure fourth order theory itself rather than through an interplay
between the second and fourth order theories. However, whether pure
fourth order theories are to possess energy eigenstates with negative
norm is something which has to be investigated in and of itself, and
we are not aware of any demonstration in the literature that pure
fourth order theories  actually do possess negative norm energy
eigenstates.

\bibitem{Dirac1964} P. A. M. Dirac, Lectures on Quantum Mechanics,
Belfer Graduate School of Science, Yeshiva University, New York (1964).
\bibitem{Pais1950} A. Pais and G. E. Uhlenbeck, Phys. Rev.  {\bf 79},
145 (1950).

\bibitem{footnote1} The need to properly take constraints into
account was first noted by us in an earlier version of this paper (P.
D. Mannheim and A. Davidson, Fourth order theories without ghosts,
hep-th/0001115, January 2000 (unpublished)) where the Dirac
quantization was first presented, and also by Hawking and Hertog (S.
W. Hawking and T. Hertog, Phys. Rev. D {\bf 65}, 103515 (2002)) who
quantized the Pais-Uhlenbeck oscillator by Feynman path integration
techniques.



\bibitem{footnote2} Indication  of the singular nature of this limit
is already seen at the classical level, with the equal frequency
solution $q(t)=c_1e^{-i\omega t}+c_2te^{-i\omega t} +{\rm c.c.}$ to Eq.
(\ref{2}) (where $\omega_1=\omega_2=\omega$) having a time behavior
quite different from that of the unequal frequency solution where
$q(t)=a_1e^{-i\omega_1t}+a_2e^{-i\omega_2t}+{\rm c.c.}$. 


\bibitem{Ostrogradski1850} M. Ostrogradski, "Memoires sur les 
equations differentielles relatives au probleme des isoperimetres",
Mem. Ac. St. Petersbourg, VI {\bf 4}, 385 (1850).

\bibitem{Bak1994} D. Bak, D. Cangemi and R. Jackiw, Phys. Rev. D {\bf
49}, 5173 (1994).

\bibitem{Boulware1983} D. G. Boulware, G. T. Horowitz, and A.
Strominger, Phys. Rev. Letts. {\bf 50}, 1726 (1983).

\bibitem{footnote2a} While asymptotic flatness is natural to second order
theories where the exterior solution to the second order Poisson equation
$\nabla^2\phi(r)=g(r)$ is given by the falling potential
$\phi(r>R)=-\alpha/r$ where
$\alpha=\int_0^Rdr^{\prime}g(r^{\prime})r^{\prime 2}$, for the fourth
order $\nabla^4\phi(r)=g(r)$ the solution is instead
given by (P. D. Mannheim and D. Kazanas, Gen. Relativ. Gravit. {\bf 26},
337 (1994)) the rising potential $\phi(r>R)=-\beta/r+\gamma r$ where
$\beta=\int_0^Rdr^{\prime}g(r^{\prime})r^{\prime 4}/6$,
$\gamma=-\int_0^Rdr^{\prime}g(r^{\prime})r^{\prime 2}/2$. Since there is
no reason to expect $\int_0^Rdr^{\prime}g(r^{\prime})r^{\prime 2}$
to vanish for an arbitrary source, even in the static
sector asymptotic flatness would not appear to be an appropriate
restriction to place on fourth order theories.


\bibitem{footnote3} In general, equal time commutators such as
$[q(t),p_q(t)]=i$ can be satisfied at all times by an
$[a,a^{\dagger}]=1$ commutator defined via $q(t)=ae^{if(t)}/2^{1/2}+{\rm
H.c.}$, $p_q(t)=ia^{\dagger}e^{-if(t)}/2^{1/2}+{\rm H.c.}$ with the
function
$f(t)$ being arbitrary. The occupation number Fock space can thus be
defined independent of the structure of $H$, and even independent of 
any interaction terms that might also be added on to the free
particle $H$. Thus in general the dimensionality of the Fock space
basis need not be the same as that of the eigenspectrum of $H$, and
their basis states need not be in one to one correspondence.

\bibitem{footnote4} Identifying $a_2^{\dagger}$ as the annihilator of
the Fock vacuum would eliminate negative norm states but would leave
$H$ without any lower bound on the ground state energy.

\bibitem{footnote5} In their study of the Pais-Uhlenbeck
oscillator, Hawking and Hertog actually argue that interactions do not
in fact lead to any such transitions, thus making the $\omega_1 \neq
\omega_2$ theory viable.

\bibitem{footnote5a} It is $b^{\dagger}$ itself which acts as a
ladder operator, though $a^{\dagger}$ does not.


\bibitem{footnote6} The unequal frequency theory states $|+1\rangle
\pm |-1 \rangle$ both have zero norm.

\bibitem{footnote7} The zero norm states 
$(b^{\dagger})^2|\Omega\rangle$
and $(a^{\dagger})^2|\Omega\rangle$ can be combined into positive and
negative norm states while the state $a^{\dagger}
b^{\dagger}|\Omega\rangle/\hat{\mu}$ has norm plus one.

\bibitem{footnote8} A further unusual property of these states is that
the one particle matrix element $\langle \Omega| q(t) b^{\dagger}|\Omega
\rangle$ is given by $\hat{\mu}e^{-i\omega t}$ rather than by the
coefficient of the $b$ field operator in Eq. (\ref{28}), with there
thus being a mismatch between the second quantized states and the first
quantized wave functions.

\bibitem{footnote8a} While we ordinarily use Hermitian matrices in quantum
mechanics since a Hermitian matrix necessarily has real eigenvalues,
there is, nonetheless, no converse theorem which would require the
eigenvalues of a non-Hermitian matrix to be complex. Rather, hermiticity is
only sufficient to secure real eigenvalues but not necessary, as the matrix
$M$ of Eq. (45) for instance directly demonstrates. Indeed, if we take a
diagonal matrix $M_1$ and add on to it a second matrix $M_2$ whose 
diagonal entries are all zero and whose only non-zero entries are all
located on only one of the two sides of the leading diagonal, then
no matter what specific values these non-zero entries actually
take, the secular equations for $M_1$ and
$M_3=M_1+M_2$ will be absolutely identical, with $M_2$  not contributing
to $|M_3-\lambda I|=|M_1-\lambda I|$ at all. As such, we can think of
matrices such as $M_2$ as being divisors of zero, so that just as the
addition of zero to an ordinary number does not change its value, the
addition of a divisor of zero such as $M_2$ to a diagonal matrix does not
change its eigenvalues. While there is no change in the eigenvalues,
there can nonetheless still be a change in the eigenvectors, with there
being a possible reduction in the number of eigenvectors in certain
cases. It is thus possible to have a non-Hermitian $N$-dimensional
Hamiltonian with $N$ real eigenvalues but less than $N$ eigenvectors.
Apart from our interest here in the application of this phenomenon to the
equal frequency fourth order oscillator, in passing we note that if
applied to neutrinos, it could result in the non-observability of any
right-handed or sterile neutrinos which might accompany the ordinary
left-handed ones.

\bibitem{footnote8b} This reduction in the number of energy eigenstates in
the $\epsilon\rightarrow 0$ limit is a quantum-mechanical reflection of
the fact that the equal frequency classical energy $H_{\rm STAT}
(\omega_1=\omega_2)$ given in Eq. (\ref{19d}) becomes
completely independent of the coefficient $c_1$ of the classical solution
$q(t)=c_1e^{-i\omega t}+c_2te^{-i\omega t}+{\rm c.c.}$ when the 
$c_2$ coefficient is set equal to zero, with
$c_1$ thus not representing a dynamical degree of freedom whose quantum
analog could materialize as an on shell state.

\bibitem{footnote8c} The authors are indebted to
Dr. A. Smilga for a very helpful comment in this regard, and also would
like to thank him for informing them of his own recent work on ghost
issues in fourth order theories -- A. V. Smilga, Benign vs malicious
ghosts in  higher-derivative theories, hep-th/0407231 (2004).


\bibitem{footnote8d} For further discussion of such
quantum-mechanical issues, see C. M. Bender, D. C. Brody and H. F. Jones,
Am. J. Phys. {\bf 71}, 1095 (2003). 

\bibitem{footnote9} A model in which a field has a positive frequency
part which annihilates the vacuum strongly and a negative frequency
part which annihilates the vacuum weakly in a way such that only
certain multiparticle states survive would appear to be a  possible
candidate mechanism for quark confinement, with quarks themselves
then only existing off shell.

\bibitem{footnote10} Within such pure fourth order theories one
of particular interest is the fourth order conformal gravity
theory, a theory which has been found capable (P. D. Mannheim,
Astrophys. J.  {\bf 561}, 1 (2001)) of readily resolving the dark
matter and dark energy problems which currently challenge the standard
second order Newton-Einstein gravitational theory. The conformal theory
is based on the imposition of the local Weyl conformal invariance
$g_{\mu\nu}(x) \rightarrow e^{2\alpha(x)} g_{\mu \nu}(x)$, and thus has a
unique  gravitational action of the form $I_W=-\alpha_g \int d^4x (-g)^{1/2}
C_{\lambda\mu\nu\kappa}  C^{\lambda\mu\nu\kappa}$ where
$C^{\lambda\mu\nu\kappa}$ is the conformal Weyl tensor and $\alpha_g$
is a dimensionless coupling constant. With the general conformal
gravity rank two gravitational tensor $-(-g)^{-1/2}\delta I_W / \delta
g_{\mu\nu}= 2\alpha_g W^{\mu\nu}$ reducing to $W^{\mu\nu}=\Pi^{\mu\rho}
\Pi ^{\nu\sigma}K_{\rho \sigma}/2- \Pi^{\mu \nu} \Pi ^{\rho
\sigma}K_{\rho\sigma}/6$ in a $g_{\mu\nu}=\eta_{\mu\nu}
+h_{\mu\nu}$ linearization around flat spacetime (here $K^{\mu
\nu}=h^{\mu \nu}-
\eta^{\mu \nu} h^{\alpha}_{\phantom{\alpha}\alpha}/4$ and $\Pi^{\mu \nu}
=\eta^{\mu \nu} \partial^{\alpha}\partial_{\alpha}-
\partial^{\mu}\partial^{\nu}$), we find in the
conformal gauge $\partial_{\nu}g^{\mu \nu}
-g^{\mu\sigma}g_{\nu\rho}\partial_{\sigma}g^{\nu 
\rho}/4=0$ (viz. a gauge condition which is left invariant under 
$g_{\mu \nu}(x)\rightarrow e^{2\alpha(x)}g_{\mu \nu}(x)$) that the
source-free region gravitational fluctuation wave equation then reduces
to $\alpha_g (\partial_0^2-\nabla^2)^2K^{\mu\nu}=0$. With this equation
of motion being decoupled in its tensor indices, we see that each tensor
component precisely obeys none other than the $M^2=0$ limit of Eq.
(\ref{1}). Since the conformal gravity theory is power counting
renormalizable (due to the dimensionlessness of $\alpha_g$), it would
thus be of interest to see if the structure we have found for the
equal frequency Pais-Uhlenbeck theory might carry over to the conformal
gravity theory, since it might then be possible to construct a
fully renormalizable, fully unitary gravitational theory in four spacetime
dimensions, one which despite its fourth order equation of motion,
would nonetheless only possess one on-shell graviton and not two.  


\end{thebibliography}
\end{document}